\documentclass[a4paper,11pt]{article}
\pdfoutput=1 

\usepackage{jheppub} 

\usepackage[T1]{fontenc} 

\newcommand{\Z}{\mathbb Z}

\newcommand{\dif}{\mbox{d}}
\newcommand{\CP}{{\cal P}}
\newcommand{\BMS}{\text{BMS}}
\usepackage{xcolor}
\usepackage{url}

\title{\boldmath  Particle realization of Bondi-Metzner-Sachs symmetry in $2+1$ space-time}

\author[a]{Carles Batlle,}
\author[b,c]{V\'{i}ctor Campello,}
\author[d]{Joaquim Gomis}

\affiliation[a]{Departament de Matem\`atiques and IOC, 
	Universitat Polit\`ecnica de Catalunya\\
	EPSEVG, Av. V. Balaguer 1, E-08800 Vilanova i la Geltr\'u, Spain}
\affiliation[b]{Universidad Internacional de Valencia, Carrer Pintor Sorolla, 21, E-46002 Val\`{e}ncia, Spain}
\affiliation[c]{Departament de Matem\`atiques, 
	Universitat Polit\`ecnica de Catalunya,\\
	Campus Diagonal Sud, Carrer de Pau Gargallo, 14, 08028 Barcelona, Spain}
\affiliation[d]{Departament de F\'isica Qu\`{a}ntica i Astrof\'isica and Institut de Ci\`{e}ncies del Cosmos (ICCUB), Universitat de Barcelona, Mart\'i i Franqu\`{e}s 1, E-08028 Barcelona, Spain}

\emailAdd{carles.batlle@upc.edu}
\emailAdd{vicmancr@gmail.com}
\emailAdd{joaquim.gomis@ub.edu}

\abstract{We construct a Lorentz invariant massive particle model in (2+1) space-time with an enlarged set of symmetries which includes Bondi-Metzner-Sachs (BMS) translations (supertranslations), using the non-linear realization framework.  The Hamiltonian formalism for the resulting Lagrangian is constructed, and the infinite phase-space constraints and the set of gauge transformations are analysed. We also compute the massless limit of the theory in phase-space. After eliminating the gauge degrees of freedom, the physical reduced space is left only with the degrees of freedom of a standard Poincar\'e particle but with a residual set of symmetries that we prove to be BMS. A similar result for the massless limit, including in this case superrotations, is pointed out.

}

\begin{document} 
	\maketitle
	\flushbottom

\section{Introduction}
\label{secIntro}

Bondi-Metzner-Sachs symmetry \cite{Bondi:1962px,Sachs:1962wk}, which originated as an extension of Poincar\'{e} symmetry for asymptotically flat space-times, has received a lot of renewed attention in the last 15 years. One of the interest is to deduce Weinberg's soft graviton theorems \cite{Weinberg:1965nx} as the Ward identities of BMS supertranslations \cite{He:2014laa,He:2014cra,Campiglia:2015qka,Kapec:2015ena}. An overview of recent developments can be found in \cite{Strominger:2017zoo}. For the relation of BMS symmetry with celestial holography see, for example, \cite{Pasterski:2021rjz,Raclariu:2021zjz} 

The BMS group, which is the semi-direct product of Lorentz and supertraslations (including ordinary space-time translations) was extended in \cite{Barnich:2009se,Barnich:2011ct,Barnich:2010eb} with the inclusion of superrotations, obtaining the extended BMS group, given by the semidirect product of { Lorentz and } superrotations and supertranslations.

A canonical realization of the BMS group was constructed using the Fourier modes of a free massive Klein-Gordon (KG) field in $2+1$. In the case of a massless KG field, one can construct an extended BMS symmetry which includes superrotations \cite{Batlle:2017llu}. This approach was pioneered in \cite{Longhi:1997zt}. In this context, BMS symmetries are not associated with asymptotically flat space-times. Instead, they appear as a generalization of the components $p^\mu$ of ordinary momenta, by noting that the supertranslations obey the Beltrami differential equation of the hyperbolic space $H_{2+1}$ or alternatively as the differential equation associated with one of the Lorentz Casimirs, $C=J^2-\vec K^2$, where $J$ is the angular momentum and $\vec K$ are the two-dimensional boosts. The massless case is obtained from a suitable limit of the Beltrami equation or through the Lorentz Casimir $C$ (see references above and also \cite{Delmastro:2017erq}).

With the idea of further exploring BMS symmetry, we propose a massive particle realization 
in $2+1$ dimensions, based on non-linear realizations of symmetry algebras (see \cite{Gomis:2006xw,Bergshoeff:2022eog} and references therein). The particle Lagrangian is constructed in an infinite-dimensional space, generalizing the Minkowski space, that we call BMS space. The infinite number of coordinates are associated with the supertranslations, and we also use two Goldstone coordinates associated with the two broken boost generators.

	We want to construct a Lagrangian in terms of the pull-back of the Maurer–Cartan forms subject to two conditions: it should have the lowest possible number of derivatives and it should be invariant under the unbroken $SO(2)$ subgroup of the Lorentz group in $2+1$, which is the symmetry that is preserved by the presence of a massive spinless point particle. One can therefore choose any component of the Maurer–Cartan form which is invariant under rotations and we take the component $\Omega_{P_0}$ along the generator  of time translations $P_0$.
	In $2+1$ dimensions one can also consider the term proportional to the generator of rotations, which can be used to construct models of particles with spin (see, for example, \cite{PLYUSHCHAY199154,Mezincescu_2010, Gomis_2013, Batlle:2014sca}), but in this paper we are only interested in the spinless case.
	 The Lagrangian is then the pull-back of $\Omega_{P_0}$ to the world-line of the particle, up to a multiplicative constant that makes the role of the mass. If one does not consider the super-translation degrees of freedom, this method produces the Lagrangian of a standard spinless massive particle in canonical form. More details of the method can be found, for instance, in Section 5.1 of \cite{Bergshoeff:2022eog}.

 The Hamiltonian formalism for the resulting Lagrangian is constructed, and the infinite phase-space first-class constraints and set of gauge transformations are analysed. We also compute the massless limit of the model. We obtain the physical reduced space after eliminating the gauge degrees of freedom by introducing an infinite set of gauge fixing constraints. This space is left only with the degrees of freedom of a standard Poincar\'e particle. Since in the gauge fixing procedure the rigid symmetries are maintained we prove that the ordinary relativistic massive Poincar\'e  particle is invariant under a realization of the BMS symmetry that we construct using the compensating gauge transformations, which are necessary to preserve the gauge fixing under supertranslations. In the massless case, we further find an infinite set of superrotations that are symmetries of the massless relativistic particle. It turns out that the
 	 infinite set of BMS coordinates associated with the supertraslations are not physical because they can always be gauged away, and therefore the model does not have  the so-called soft BMS modes. These results agree with the fact that the quadratic Casimir of BMS algebra coincides with the quadratic Casimir of the Poincar\'e algebra.

A different approach to the definition of BMS particles in $2+1$ dimensions is based on the coadjoint orbit approach \cite{Oblak:2015sea,Campoleoni:2016vsh,Oblak:2016eij,Ciambelli:2022cfr}. However, as far as we know,  no particle action has been constructed in the literature using this approach. For the relation among the non-linear realization framework and the coadjoint orbit method, see for example
\cite{Bergshoeff:2022eog}.

The paper is organized as follows. Section \ref{secNLR} we derive the BMS particle Lagrangian in $2+1$ space-time using the
non-linear realization approach. The canonical analysis of this Lagrangian is given in Section \ref{secCA}, including the discussion of the constraints, the reduced phase space and the gauge transformations induced by the first-class constraints. Section \ref{secGBMS} presents the generators that realize the Poincar\'{e} symmetry in BMS coordinates, and shows that the theory is indeed invariant under them. The massless limit of the theory is computed in Section \ref{secM}, and it is seen that the set of symmetry generators is extended to include superrotations. The gauge fixing of the theory is presented in Section \ref{secGF}, and the physical degrees of freedom of the BMS particle are determined. Our results are discussed in Section \ref{secCon}, and some possible extensions and connections with other approaches are also considered. 
Detailed proofs of some of the results are presented in Appendixes (\ref{appA}), (\ref{appVL}) and (\ref{appM}), and Appendix (\ref{secC}) contains a discussion of the 
quadratic Casimir  in BMS space.

\section{Non-linear realization of the BMS algebra in $2+1$ dimensions}
\label{secNLR}
The extended BMS algebra in $2+1$ dimensions \cite{Barnich:2011ct} without central extensions is given by 

\begin{eqnarray}
   [L_n,P_m] &=& i (n-m) P_{n+m},\nonumber\\ 
  \left[P_n,P_m\right] &=& 0,\nonumber\\    
  \left[L_n,L_m\right] &=& i (n-m) L_{n+m},
  \label{fullBMS}
\end{eqnarray}
with $m,n\in \Z$.
We are interested in the subalgebra $BMS_3$ formed by the Lorentz generators $L_0$, $L_{\pm1}$ and the supertranslations $P_n$:
\begin{align}
&	\left[L_{1},L_{-1}\right] = 2iL_0,\quad
		\left[L_{1},L_{0}\right] = iL_1,\quad
			\left[L_{-1},L_{0}\right] = -iL_{-1},\label{LL}\\
&	\left[L_{-1},P_m\right] = -i (m+1) P_{m-1},\quad
	 \left[L_{0},P_m\right] = -i m P_{m},\quad
	 \left[L_{1},P_m\right] = -i (m-1) P_{m+1},\label{LP}\\
&	\left[P_n,P_m\right] = 0. \label{PP}   	 
\end{align}

In order to construct a massive BMS particle  we should consider the coset $BMS_3/SO(2)$, locally given by
\begin{equation}
g(\{x\},u,v)= g_0(\{x\})e^{i L_{-1}v} e^{iL_1 u}=g_0(\{x\}) U(u,v),	
\label{g1}	
\end{equation}
with
\begin{equation}
g_0(\{x\})=\prod_{n\in\Z} e^{i P_n x^n}.
\label{g0}
\end{equation}
Here $U(u,v)$ is a boost transformation generated by $L_1$, $L_{-1}$ and parametrized by the Goldstone coordinates
$u, v$
and $\{x^n\}_{n\in\Z}$ are the BMS coordinates (with $x^0$, $x^{\pm 1}$ related to ordinary $2+1$  space-time). 
The BMS coordinates are complex, with $x^n$ and $x^{-n}$ complex conjugate of each other.

The Maurer-Cartan form associated to $g$ is
\begin{align}
	\Omega(g) &= -i g^{-1}\dif g
	=-iU^{-1} g_0^{-1} \dif g_0\, U -i U^{-1} \dif U \label{MCg0}\\
	&=\sum_{n\in\Z} \dif x^n\, U^{-1}P_n U -i U^{-1}\dif U.
	\label{MCg}
\end{align}

In the spirit of obtaining spinless particle actions, {see for example}  \cite{Gomis:2006xw,Bergshoeff:2022eog}, we are only interested in the terms $\Omega_{P_0}$ of $\Omega$ proportional to $P_0$, which can only come from $U^{-1}P_n U$. The detailed computation is presented in Appendix \ref{appA}, and the result is
\begin{equation}
\Omega_{P_{0}}= 	\dif x^0 (1-2uv)
	+ \sum_{n=1}^\infty  \dif x^n  
	(-1)^n v^n (n+1- 2 uv)
	+ \dif x^{-1} 2u+    \sum_{n=2}^\infty 
	\dif x^{-n} 
	u^n  \frac{n+1-2uv}{(1-uv)^{n}}.
	\label{BMS_P0}
\end{equation}


Following the standard procedure, we integrate the pullback of $\Omega_{P_{0}}$ to the world-line of the particle
\begin{eqnarray}
	\lefteqn{
		S[\{x\},u,v] = -\mu \int\dif\tau (
		\dot x^0 (1-2uv) - 2 \dot x^1 v (1-uv) + 2 \dot x^{-1} u \nonumber}\\
	&+& \sum_{n=2}^\infty \dot x^n (-1)^n v^n (n+1-2uv) + \sum_{n=2}^\infty \dot x^{-n} u^n \frac{n+1-2uv}{(1-uv)^n}
	)\nonumber\\
	&=& 
	-\mu \int\!\!\dif\tau\! \left(
	\sum_{n=0}^\infty \dot x^n (-1)^n v^n (n+1-2uv) + \sum_{n=1}^\infty \dot x^{-n} u^n \frac{n+1-2uv}{(1-uv)^n}
	\right),
	\label{BMS_act}
\end{eqnarray}
and define the BMS particle Lagrangian
\begin{equation}
	{\cal L} = -\mu \left(
	\sum_{n=0}^\infty \dot x^n (-1)^n v^n (n+1-2uv) + \sum_{n=1}^\infty \dot x^{-n} u^n \frac{n+1-2uv}{(1-uv)^n}
	\right).
	\label{BMS_lag}
\end{equation}

The contribution to (\ref{BMS_act}) of the ordinary space-time coordinates, \textit{i.e.}  $x^0$, $x^{\pm 1}$, is
\begin{equation}
S_{0} = -\mu \int\dif \tau \left(
\dot x^0 (1-2uv) +  2\dot x^1 (-v+ v^2 u)+ 2 \dot x^{-1} u
\right) \equiv \int\dif\tau {\cal L}_{0}.
\label{S0}
\end{equation}
This action corresponds to an ordinary spinless relativistic particle in flat (2+1) Minkowski space-time, as can be seen by computing the  momenta
\begin{eqnarray}
p_0 &=& \frac{\partial  {\cal L}_{0} }{\partial \dot x^0} =-\mu( 1-2uv),\label{p0uv}\\
p_1 &=& \frac{\partial  {\cal L}_{0} }{\partial \dot x^1} = -2\mu (-v+v^2 u),\label{p1uv}\\
p_{-1} &=& \frac{\partial  {\cal L}_{0} }{\partial \dot x^{-1}} = -2\mu u,\label{q1uv}
\end{eqnarray}
and checking the mass-shell condition
\begin{equation}
-p_0^2 + p_1 p_{-1} = -\mu^2.
\end{equation}
Actually, if one computes the   EOM  given by (\ref{S0}) for the boost variables,
\begin{eqnarray}
- v \dot x^0 + v^2 \dot x^1 + \dot x^{-1} &=& 0,\\
-u \dot x^0 -\dot x^1 + 2uv \dot x^1 &=& 0,
\end{eqnarray}
solves them for $u$, $v$,
\begin{equation}
u =\frac{\dot x^1}{\pm\sqrt{(\dot x^0)^2 - 4 \dot x^1\dot x^{-1}}},\quad
v= \frac{\dot x^0 \pm        \sqrt{(\dot x^0)^2 - 4 \dot x^1\dot x^{-1}}          }{2\dot x^1},
\end{equation}
makes the change of space variables from the complex ones $x^{\pm 1}$  to the real ones $x_1$, $x_2$,  given by
\begin{equation}
x^{\pm 1} = \frac{1}{2} (x_1\pm i x_2),
\label{realx}
\end{equation}
and substitutes the resulting expressions for $u$, $v$ in (\ref{S0}), one gets the ordinary space-time action with Lagrangian
\begin{equation}
{\cal L}_{0}^* = \mp \mu \sqrt{(\dot x^0)^2 - (\dot x_1)^2-(\dot x_2)^2}.
\label{L0}
\end{equation}

\section{Canonical analysis of the BMS particle action}
\label{secCA}
In order to understand the structure of the BMS Lagrangian
(\ref{BMS_lag}) we perform the Hamiltonian analysis. The momenta are given by
\begin{eqnarray}
\pi_u &=&\frac{\partial{\cal L}}{\partial \dot u} = 0,\\
\pi_v &=&\frac{\partial{\cal L}}{\partial \dot v} = 0,\\
p_n &=& \frac{\partial{\cal L}}{\partial \dot{x}^n}= -\mu (-1)^n v^n (n+1-2uv),\ \ n=0,1,2,\ldots, \label{pn}\\
\bar{p}_n &=&  \frac{\partial{\cal L}}{\partial \dot{x}^{-n}}=- \mu u^n \frac{n+1-2uv}{(1-uv)^n},\ \ n=1,2,\ldots. \label{bpn}
\end{eqnarray}
Notice that, since the $x^{-n}$ are complex conjugates of the $x^{n}$, then the fact that ${\cal L}$ is real implies that $p_n$ and $\bar{p}_n$ are also complex conjugates of each other.
 
From the expressions of the momenta one gets the set of primary constraints $\pi_u=0$, $\pi_v=0$ and $\phi_n=0$, $\bar{\phi}_n=0$, with
\begin{eqnarray}
\phi_n &:=& p_n+\mu (-1)^n v^n (n+1-2uv),\ \ n=0,1,2,\ldots\\
\bar{\phi}_n &:=& \bar{p}_n+ \mu u^n \frac{n+1-2uv}{(1-uv)^n},\ \ n=1,2,\ldots
\end{eqnarray}
The non-zero Poisson brackets between these constraints are
\begin{eqnarray}
\{  \phi_n, \pi_u \} &=& - 2 \mu (-1)^n v^{n+1},\ n=0,1,2,\ldots\\
\{  \phi_n, \pi_v \} &=& \mu (-1)^n  v^{n-1}( n(n+1) - 2 (n+1)uv), \ n=0,1,2,\ldots\\
\{     \bar{\phi}_n , \pi_u      \} &=& \frac{\mu u^{n-1}}{(1-uv)^{n+1}} (2u^2v^2 - 2nuv +n^2 -2uv +n), \ n=1,2,\ldots\\
\{     \bar{\phi}_n , \pi_v     \} &=& \frac{\mu u^{n+1}}{(1-uv)^{n+1}} (n-1) (-2uv +n +2), \ n=1,2,\ldots
\end{eqnarray}

Since the Lagrangian is homogeneous of degree one in the velocities, the canonical Hamiltonian is identically zero and one must consider the
Dirac Hamiltonian
\begin{equation}
H_D = \lambda_u \pi_u + \lambda_v \pi_v + \sum_{n=0}^\infty \lambda_n \phi_n +\sum_{n=1}^\infty \bar{\lambda}_n \bar{\phi}_n,
\end{equation}
where the $\lambda$ are arbitrary functions.

If we order the constraints as $(\pi_u, \pi_v, \phi_1, \bar{\phi}_1, \phi_2, \bar{\phi}_2,\ldots)$, the infinite-dimensional matrix of Poisson brackets between them has the form
\begin{equation}
	M_\infty = \left(
	\begin{array}{ccccc}
		 0 & A_1 & A_2 & A_3 & \cdots \\
		 -A_1^T & 0 & 0 & 0 &  \cdots\\
		 - A_2^T & 0 & 0 & 0 &  \cdots\\
		 - A_3^T & 0 & 0 & 0 &  \cdots\\
		 \vdots & \vdots & \vdots & \vdots &  \ddots
	\end{array}
	\right),
\end{equation}
where all the entries are $2\times 2$ blocks and
\begin{equation}
	A_i = \left(
	\begin{array}{cc}
		\{\pi_u,\phi_i\} & \{\pi_u,\bar{\phi}_i\}\\
		\{\pi_v,\phi_i\} & \{\pi_v,\bar{\phi}_i\}\\
	\end{array}	
	\right),\quad  i=1,2,3,\ldots
\end{equation}
Since, for instance,
\begin{equation}
	A_1 = 2\mu\left(
	\begin{array}{cc}
		-v^2 & -1\\
		1-2uv & 0
	\end{array}	
	\right), 
\end{equation}
has non-zero determinant, provided that $1-2uv\neq 0$, it turns out that all the (infinite dimensional) columns to the right of $A_1$ can be expressed as linear combinations of the columns which contain $A_1$.  Hence, considering also the first two columns of $M_\infty$, one can show that $M_\infty$ has a rank equal to $4$. This means that, at most, we can select $4$ second-class constraints, including necessarily $\pi_u$ and $\pi_v$, plus another two which allow us to eliminate $u$ and $v$ in terms of two of the momenta $p_i$ and $\bar{p}_i$. Also, notice that the number of first class constraints is infinite.

Notice that, although $u$, $v$ can be eliminated from any two of the $\phi$, $\bar{\phi}$, it is convenient to select $\phi_1$, $\bar{\phi}_1$, as it was done for the case of the pure Poincar\'{e} particle. The four selected constraints
\begin{equation}
\pi_u, \ \ \pi_v, \ \ \phi_1, \ \ \bar{\phi}_1
\end{equation}
are second class, with the Poisson bracket matrix
\begin{equation}
M=
\left(
\begin{array}{cc}
	0 & A_1 \\
	-A_1^T & 0 \\
\end{array}
\right)
=2\mu \begin{pmatrix}
0 & 0 & -v^2 & -1 \\ 0 & 0 & 1-2uv & 0 \\ v^2 & -1+2uv & 0 & 0 \\ 1 & 0 & 0 & 0 
\end{pmatrix},
\end{equation}
which has determinant $16 \mu^4 (1-2uv)^2$ and inverse 
\begin{equation}
M^{-1}=\frac{1}{2\mu} \begin{pmatrix}
0 & 0 & 0 & 1 \\ 0 & 0 & -\frac{1}{1-2uv} & \frac{v^2}{1-2uv} \\ 0 & \frac{1}{1-2uv} & 0 & 0 \\ -1 & -\frac{v^2}{1-2uv} & 0 & 0 
\end{pmatrix}.
\end{equation}
From this, one can define the Dirac bracket
\begin{equation}
\{ A,B \}_D = \{ A,B \}-\{ A, \Psi_i\}M^{-1}_{ij}\{\Psi_j,B\},
\end{equation}
where $\Psi_j\in\{       \pi_u, \pi_v,  \phi_1,  \bar{\phi}_1   \}$. 
 
If one demands the stability of the primary second-class constraints, \textit{i.e.} 
\begin{equation}
\dot\Psi_i = \{\Psi_i,H_D\} \stackrel{{\cal M}}{=}0,
\end{equation}
where ${\cal M}$ is the submanifold defined by the constraints $\Psi_i$, one can determine the values of the four arbitrary functions $\lambda_u$, $\lambda_v$, $\lambda_1$, $\bar{\lambda}_1$. The result is that $\lambda_u=\lambda_v=0$, while $\lambda_1$ and $\bar{\lambda}_1$ are quite involved functions of  all the other $\lambda$. However, in the reduced space ${\cal M}$ we can set $\phi_1=\bar{\phi}_1=0$, and the reduced Dirac Hamiltonian is
\begin{equation}\label{redH}
H_{\cal M} = \lambda_0 \phi_0 + \sum_{n=2}^\infty (\lambda_n \phi_n + \bar{\lambda}_n \bar{\phi_n}),
\end{equation}
where all the constraints are assumed to be computed on ${\cal M}$. Using $\phi_1=0$ and $\bar{\phi}_1=0$ to effectively eliminate $u$ and $v$ in terms of $p_1$ and $\bar{p}_{1}$ one has
\begin{eqnarray}
u &=& -\frac{1}{2\mu}\bar{p}_1,\label{up1}\\
v &=& -\frac{\mu}{\bar{p}_1}\pm \frac{1}{\bar{p}_1} \sqrt{\mu^2+p_1\bar{p}_1},\label{vp1}
\end{eqnarray}
from which it also follows that
\begin{equation}
uv = \frac{1}{2} \mp \frac{1}{2\mu}\sqrt{\mu^2+p_1\bar{p}_1}.
\end{equation}
Then, on the reduced space,
\begin{equation}\label{redphi0}
\phi_0 = p_0 \pm  \sqrt{\mu^2+p_1\bar{p}_1},
\end{equation}
  and
\begin{eqnarray}
\phi_n &=& p_n + \mu (-1)^n \left( 
 -\frac{\mu}{\bar{p}_1}\pm \frac{1}{\bar{p}_1} \sqrt{\mu^2+p_1\bar{p}_1}
\right) ^n \left(
n \pm \frac{1}{\mu}\sqrt{\mu^2+p_1\bar{p}_1}
\right), \label{redphi1}\\
\bar{\phi}_n &=&\bar{p}_n +\mu (-1)^n \bar{p}_1^n 
\frac{
n \pm \frac{1}{\mu}\sqrt{\mu^2+p_1\bar{p}_1}
}{
\left(
\mu \pm \sqrt{\mu^2+p_1\bar{p}_1}
\right)^n
}\nonumber\\
&=& 
\bar{p}_n + \mu (-1)^n \left( 
\frac{\mu}{\bar{p}_1}\pm \frac{1}{\bar{p}_1} \sqrt{\mu^2+p_1\bar{p}_1}
\right)^{-n} \left(
n \pm \frac{1}{\mu}\sqrt{\mu^2+p_1\bar{p}_1}
\right),\label{redphi2}
\end{eqnarray}
for $n\geq 2$ (as a check, these expressions become identities for $n=1$). The case $n=0$ of (\ref{redphi0}) can also be included in either (\ref{redphi1}) or (\ref{redphi2}). 
Notice that the mass-shell condition can be recovered by taking the square of $\phi_0$. This is inherent to the nonlinear realization formalism, since the momenta appear always linearly in the canonical constraints (see \cite{Gomis_2013} for a discussion of this mechanism in the case of $p$-branes).

The constraints can also be written as
\begin{align}
	\phi_n &= p_n +  \frac{\mu}{\bar{p}_1^n} f_n^\pm(p_1\bar{p}_1),\label{rphi1}\\
	\bar{\phi}_n &= \bar{p}_n + \mu \bar{p}_1^n g_n^\pm(p_1\bar{p}_1),\label{rphi2}
\end{align}
for $n=0,2,3,\ldots$, where
\begin{align}
	f_n^\pm(x) & = \left(\mu\mp\sqrt{\mu^2+x}\right)^n \left(n\pm \frac{1}{\mu}\sqrt{\mu^2+x}\right),\label{deff}\\
	g_n^\pm(x) & = \left(-\mu\mp\sqrt{\mu^2+x}\right)^{-n} \left(n\pm \frac{1}{\mu}\sqrt{\mu^2+x}\right),
	\label{defg}
\end{align}
which satisfy, for $n\geq 1$,
\begin{align}
\frac{\dif}{\dif x}	f_n^\pm(x) & = \mp \frac{n+1}{2\sqrt{\mu^2+x}} f_{n-1}^\pm(x),\label{derf}\\
\frac{\dif}{\dif x}	g_n^\pm(x) & = \pm \frac{n-1}{2\sqrt{\mu^2+x}} g_{n+1}^\pm(x),
\label{derg}
\end{align}
and also the second-order recurrence relation
\begin{align}
& (n-1)f_{n+1}^\pm(x)\pm 2n \sqrt{\mu^2+x} f_n^\pm(x) + (n+1)x f_{n-1}^\pm(x)=0, \quad n\geq 1, \label{rec3f}\\
& (n+1)g_{n-1}^\pm(x)\pm 2n \sqrt{\mu^2+x} g_n^\pm(x) + (n-1)x g_{n+1}^\pm(x)=0, \quad n\geq 1.\label{rec3g}
\end{align}

The signs $\pm$ which appear in the above expressions correspond to different sheets of the constraint manifold in reduced space, parametrized by $p_1$ and $\bar{p}_{1}$. Notice that constraints $\phi_n, \bar\phi_n$ are first class. Therefore we expect that the physical degrees of freedom of the BMS particle will be finite-dimensional.

In the reduced phase space, with the boost variables $u$, $v$ eliminated in terms of $p_1$, $\bar{p}_1$, the symmetry between the momenta corresponding to coordinates with positive and negative index is restored, and also that the constraint $\bar{\phi}_n$ is the complex conjugate of $\phi_n$, thereby justifying the notation. 
It will also be convenient to introduce the functions $P_n$, $\bar{P}_n$ of the variables $P_1$, $\bar{P}_1$ defined by
\begin{align}
	P_n &= -\frac{\mu}{\bar{p}_1^n} f_n^\pm(p_1\bar{p}_1),\label{Pn}\\
	\bar{P}_n &= -\mu \bar{p}_1^n g_n^\pm(p_1\bar{p}_1),\label{bPn}
\end{align}
and, in particular,
\begin{equation}
	P_0 = \mp \sqrt{\mu^2+p_1\bar{p}_1},
\end{equation}
in terms of which the constraints are $\phi_n=p_n-P_n$, $\bar{\phi}_n = \bar{p}_n-\bar{P_n}$, $\phi_0=p_0-P_0$. Notice also that $P_1=p_1$, $\bar{P}_1 =\bar{p}_1$.

Due to the fact that $M^{-1}$ does not have contributions in the  lower-right square, the Dirac brackets of the variables $x^n$, $x^{-n}$, $p_n$, $\bar{p}_n$ do not change with respect to the Poisson ones,
\begin{eqnarray}
\{  x^n, p_m\}_D= \{  x^n, p_m\} = \delta^n_m,\\ 
\{  x^{-n}, \bar{p}_m\}_D= \{  x^{-n}, \bar{p}_m\} = \delta^n_m.
\end{eqnarray}
First of all, the $x^n$, $x^{-n}$, $p_n$, $\bar{p}_n$ have zero Poisson brackets with all the $4$ second-class constraints for $n\geq 2$, so only the cases for $n=1$ need to be discussed. Since $p_1$, $\bar{p}_1$ have also zero brackets with all the constraints, all brackets involving either of them remain also unchanged. Finally,
$$
\{  x^1, x^{-1}\}_D= 0 - \{x^1,\Psi_i\}M_{ij}^{-1}\{\Psi_j,x^{-1}\},
$$
but the only non-zero brackets of the $x^{\pm 1}$ are with $\phi_1$ and $\bar{\phi}_1$, and this selects the lower-right square of $M^{-1}$, which is identically zero. The non-trivial Dirac brackets are those involving $u$, $v$, $\pi_u$, $\pi_v$ and $x^{\pm 1}$:
\begin{align}
&	\{u,x^1\}_D = 0, \quad \{v,x^1\}_D = -\frac{1}{2\mu}\frac{1}{1-2uv},\\ 
&     	\{u,x^{-1}\}_D = \frac{1}{2\mu}, \quad \{v,x^{-1}\}_D = \frac{1}{2\mu}\frac{v^2}{1-2uv},\\
&  \{\pi_u,u\}_D = \{\pi_u,v\}_D = \{\pi_v,u\}_D = \{\pi_v,v\}_D = 0,\\
& \{\pi_u,x^1\}_D = \{\pi_u,x^{-1}\}_D = \{\pi_v,x^1\}_D = \{\pi_v,x^{-1}\}_D = 0.
\end{align}

Summing up, in the reduced phase space one has coordinates
\begin{equation}
\{
x^n,\ x^{-m}, \ p_n, \ \bar{p}_m
\}_{\stackrel{n=0,1,2,\ldots,}{m=1,2,\ldots}}
\end{equation}
with  Hamiltonian (\ref{redH}), where  the constraints are given by (\ref{redphi0}), (\ref{redphi1}), (\ref{redphi2}), and with ordinary brackets. The first class constraints $\phi_n, \bar\phi_n$ will generate infinite gauge transformations  given by the canonical generator
\begin{equation}
G= \epsilon(\tau) \phi_0 + \sum_{m\geq 2} \left(
\alpha_m(\tau) \phi_m + \beta_m(\tau) \bar{\phi}_m
\right).
\end{equation}
For instance, the re-parametrization associated to $\phi_0 = p_0 \pm\sqrt{\mu^2+p_1\bar{p}_1}$ acts only on the standard space-time coordinates, and is given by
\begin{eqnarray}
\delta x^0 &=&\epsilon(\tau) \{x^0,\phi_0\}_D = \epsilon(\tau),\\
 \delta x^1 &=&\epsilon(\tau) \{x^1,\phi_0\}_D = \pm\epsilon(\tau)\frac{\bar{p}_1}{2\sqrt{\mu^2+p_1\bar{p}_1}}=-\epsilon\frac{\bar{p}_1}{2P_0},\\
 \delta x^{-1} &=&\epsilon(\tau) \{x^{-1},\phi_0\}_D = \pm\epsilon(\tau)\frac{{p}_1}{2\sqrt{\mu^2+p_1\bar{p}_1}}=-\epsilon\frac{{p}_1}{2P_0},
\end{eqnarray} 
where the function $P_0$ of $p_1,\bar{p}_1$ has been used to rewrite the final expression. Furthermore, $\delta p_0=\delta p_1 = \delta \bar{p}_1=0$ and hence also  $\delta u=\delta v=0$.

In order to check the action of this symmetry on the Lagrangian ${\cal L}$  it suffices to consider the action on ${\cal L}_0$, since the re-parametrization acts only on $x^0, x^{\pm1}$. It is convenient to re-define the arbitrary function $\epsilon(\tau)$ as $\epsilon(\tau)/(2P_0)$, so that
\begin{equation}
	\delta x^0 = 2P_0\epsilon(\tau),\ \delta x^1 = -\epsilon(\tau)\bar{p}_1,\  \delta x^{-1} = -\epsilon(\tau) {p}_1,
\end{equation}
and also to write the Lagrangian in terms of the momenta $p_1, \bar{p}_1$,\footnote{If one uses the equations of motion for $p_1, \bar{p}_1$ to eliminate these non-dynamical variables, the result is the standard Poincaré Lagrangian in the form (\ref{L0}).}
\begin{equation}
	{\cal L}_0 = \dot{x}^0 P_0 + \dot{x}^1 p_1 + \dot{x}^{-1} \bar{p}_1.
\end{equation}
One has then, taking into account that the $p_1$, $\bar{p}_1$ do not transform under this symmetry,
\begin{align}
	\delta{\cal L}_0 &= 2 \frac{\dif}{\dif\tau}(\epsilon P_0)P_0 - 
	\frac{\dif}{\dif\tau}(\epsilon \bar{p}_1)p_1 -\frac{\dif}{\dif\tau}(\epsilon p_1)\bar{p}_1\\
	&= \epsilon (2P_0\dot{P}_0-\frac{\dif}{\dif\tau}(p_1\bar{p}_1)) +\dot\epsilon (2P_0^2-2p_1\bar{p}_1)
	= \epsilon \frac{\dif}{\dif\tau} (P_0^2-p_1\bar{p}_1) + 2\dot\epsilon (P_0^2-p_1\bar{p}_1)\\
	&= \frac{\dif}{\dif\tau}(2\mu^2\epsilon),
\end{align}
where  $P_0^2-p_1\bar{p}_1=\mu^2$ has been used. 

Similarly, one can study the transformation of the Lagrangian (\ref{BMS_lag}) under the transformation generated by $\phi_m$ for $m\geq 2$. 
We write the Lagrangian (\ref{BMS_lag}) in the notationally convenient form
\begin{equation}
	{\cal L} = \sum_{n=0}^\infty \dot{x}^n P_n + \sum_{n=1}^\infty \dot{x}^{-n} P_{-n},
	\label{BMS_lagp}
\end{equation}
where $P_{-n}=\bar{P}_n$ and all of them are functions of $p_1$, $p_{-1}=\bar{p}_1$ (or of $u$ and $v$).

 As shown in Appendix \ref{appVL}, if $G=\alpha_m(\tau)\phi_m$ one has
\begin{equation}
	\delta x^n =\{x^n,G\}_D = \begin{cases}
		\alpha_m & \text{if $n=m$},\\
		-\alpha_m (m+1) \frac{P_{m-1}}{2P_0} & \text{if $n=1$},\\
		\alpha_m (m-1) \frac{P_{m+1}}{2P_0} &\text{if $n=-1$},\\
		0 & \text{otherwise},
	\end{cases}
\end{equation} 
and then
\begin{align}
	\delta{\cal L} = \frac{\dif}{\dif\tau} \left(
	\epsilon_m (2P_0 P_m - (m+1)p_1 P_{m-1}+(m-1)\bar{p}_1 P_{m+1})
	\right),\label{VLphi}
\end{align}
where the parameter of the transformation has been written as $\alpha_m=2P_0\epsilon_m$. Notice that the function inside the total derivative depends on $p_1$, $\bar{p}_1$ and hence, through (\ref{p1uv}), (\ref{q1uv}), on the original Lagrangian variables $u,v$. Similarly, for $G=\beta_m\bar{\phi}_m$, one has
\begin{equation}
	\delta x^n =\{x^n,G\}_D = \begin{cases}
		\beta_m & \text{if $n=-m$},\\
		\beta_m (m-1) \frac{\bar{P}_{m+1}}{2P_0} & \text{if $n=1$},\\
		-\beta_m (m+1) \frac{\bar{P}_{m-1}}{2P_0} &\text{if $n=-1$},\\
		0 & \text{otherwise},
	\end{cases}
\end{equation} 
and then,
\begin{align}
	\delta{\cal L} = \frac{\dif}{\dif\tau} \left(
	\epsilon_m (2P_0 \bar{P}_m - (m+1)\bar{p}_1 \bar{P}_{m-1}+(m-1){p}_1 \bar{P}_{m+1})
	\right),\label{VLbphi}
\end{align}
with $\beta_m=2P_0\epsilon_m$. 

Notice that, in fact, these generators do not yield real transformations of ${\cal L}$, since they do not respect the fact that $x^n$ and $x^{-n}$ are complex conjugates, but this can be easily solved by working with the real generators
\begin{align}
	G_m = \epsilon_m (\phi_m + \bar{\phi}_m), \quad \epsilon_m^*=\epsilon_m,\\
	\bar{G}_m = \epsilon_m (\phi_m -\bar{\phi}_m), \quad \epsilon_m^*=-\epsilon_m,
\end{align}
for $m=2,3,\ldots$, and using (\ref{VLphi}) and (\ref{VLbphi}) to obtain the corresponding variations of the Lagrangian.

\section{Realization of Lorentz symmetry in BMS space}
\label{secGBMS}

From now on we will use the notation $p_{-n}=\bar{p}_n$ in order to obtain more compact expressions. 
The generators of the Lorentz group in physical $2+1$ space-time with coordinates $x^0$, $x_1$, $x_2$ and corresponding canonical momenta $p_0$, $P_1$, $P_2$ are given by $K_0 = x_1 P_2 - x_2 P_1$ for rotations and $K_i= x^0 P_i + x_i p_0$, $i=1,2$, for boosts. The relation with the $x^{\pm1}$ coordinates is given by $x^{\pm 1} = \frac{1}{2}(x_1\pm i x_2)$, which induces for the momenta the relation $p_{\pm1} = P_1 \mp i P_2$. Defining $J=-i K_0$, $K_\pm = K_1\mp i K_2$, one has, in terms of the coordinates $x^0$, $x^{\pm1}$ and their associated canonical momenta $p_0$, $p_{\pm 1}$,
\begin{align}
	J &= x^1 p_1 - x^{-1} p_{-1},\\
	K_+ &= x^0 p_1 + 2 p_0 x^{-1},\\
	K_- &= x^0 p_{-1} + 2 p_0 x^1,
\end{align} 
which obey the Lorentz  $SO(2,1)$ algebra $\{K_+,K_-\}=2J$, $\{J,K_+\}=K_+$, $\{J,K_-\}=-K_-$.
These generators can be extended to the BMS space with coordinates $x^{n}, p_n$, $n\in\Z$ by defining
\begin{align}
J &= \sum_{n=-\infty}^{+\infty} n x^n p_n\nonumber\\ &= \cdots - 2 x^{-2}p_{-2} \boxed{- x^{-1}p_{-1} + 0\cdot x^0 p_0 + x^1 p_1} + 2 x^2 p_2 + \cdots,\label{J}\\
K_+ &= \sum_{n=-\infty}^{+\infty} (1-n)x^n p_{n+1}\nonumber\\ &=\cdots + 3 x^{-2}p_{-1} \boxed{+ 2 x^{-1} p_0 + x^0 p_1} + 0\cdot x^1 p_2 - x^2 p_3 + \cdots,\label{K+}\\
K_- &= \sum_{n=-\infty}^{+\infty} (1+n)x^n p_{n-1}\nonumber\\ &=\cdots - x^{-2}p_{-3} + 0\cdot x^{-1} p_{-2} \boxed{+ x^0 p_{-1} + 2 x^1 p_0} +3 x^2 p_1 + \cdots,\label{K-}
\end{align}
Notice that, under complex conjugation, $(K_+)^* = K_-$, and that $J^*=-J$,  as it must be according to their definition from the real generators $K_0$, $K_1$ and $K_2$.

This set of generators, together with the supertranslation generators
\begin{equation}
	\CP_n = p_n, \label{eq:supertrans}
\end{equation}
provides a realization of BMS (Lorentz + supertranslations) in phase space, with $\{x^n,p_m\}=\delta^n_m$. Indeed, one has
\begin{align}
	& \{K_+,K_-\}=2J, \quad \{J,K_+\}=K_+,\quad \{J,K_-\}=-K_-,\label{Lorentz}\\
	& \{K_+, \CP_n\} = (1-n)\CP_{n+1},\quad \{K_-, \CP_n\} = (1+n)\CP_{n-1},\label{ladder}\\
	& \{J,\CP_n\} = n \CP_n,\quad \{\CP_n,\CP_m\}=0,\label{JPPP}
\end{align}
with $n,m\in\Z$. The connection with the abstract algebra (\ref{LL}-\ref{PP}) is made by means of the identifications $K_+\mapsto -i L_1$, $K_-\mapsto i L_{-1}$, $J\mapsto i L_0$, $\CP_n\mapsto P_n$.

The fact that the extended generators satisfy the correct algebra is not enough to state that we have a realization of BMS symmetry. Indeed, one must prove that the generators $K_+$, $K_-$, $J$, and $\CP_n$ are conserved charges of the system. In this case, one must prove that the generators have weakly zero Poisson brackets with all the first-class constraints $\phi_0$, $\phi_n$, $\bar{\phi}_n$, $n=2,3,\ldots$, appearing in the reduced Hamiltonian (\ref{redH}). Here, weakly zero means zero up to the constraints, and we will denote this by $\simeq 0$.   

Since the constraints do not depend on the coordinates, this condition is trivially satisfied by the generators of supertranslations $\CP_n$. For $J$ one has, using that $\{p_n,J\}=-np_n$, $\{p_{-n},J\} = np_{-n}$, $\{p_1p_{-1},J\}=0$,
\begin{align*}
	\{\phi_n,J\} &= \{ p_n + \mu p_{-1}^{-n} f_n^\pm (p_1p_{-1}),J\} = -n p_n + \mu  f_n^\pm (p_1p_{-1}) \{p_{-1}^{-n},J\}\nonumber\\
	&= -n p_n -n \mu f_n^\pm (p_1p_{-1}) p_{-1}^{-n-1} \{p_{-1},J\} = -n p_n -n \mu f_n^\pm (p_1p_{-1}) p_{-1}^{-n-1} p_{-1}\\ &= - n \phi_{n} \simeq 0,\\
	\{\bar{\phi}_n,J\} &= \{ p_{-n} + \mu p_{-1}^{n} g_n^\pm (p_1p_{-1}),J\} = n p_{-n} + \mu g_n^\pm (p_1p_{-1}) \{p_{-1}^{n},J\}\nonumber\\
	&= n p_{-n} + n \mu g_n^\pm (p_1p_{-1}) p_{-1}^{n-1} \{p_{-1},J\} = n p_n + n \mu g_n^\pm (p_1p_{-1}) p_{-1}^{n-1} p_{1}\\ &=  n \bar{\phi}_{n} \simeq 0.
\end{align*}


For $K_+$, using that $\{p_n,K_+\} = -(1-n) p_{n+1}$, $\{p_{-n},K_+\}=-(1+n)p_{-n+1}$, and in particular that $\{p_1,K_+\}=0$, $\{p_{-1},K_+\}=-2p_0$, 
\begin{align*}
	\{\phi_n,K_+\} &= \{p_n+\mu p_{-1}^{-n} f_n^\pm(p_1p_{-1}),K_+\}\\ 
	&= -(1-n)p_{n+1} + \mu f_n^\pm(p_1p_{-1}) \{p_{-1}^{-n},K_+\} + \mu p_{-1}^{-n} \{f_n^\pm(p_1p_{-1}),K_+\}\\
	&=-(1-n)p_{n+1} + \mu f_n^\pm(p_1p_{-1}) (-n p_{-1}^{-n-1} (-2p_0) ) + \mu p_{-1}^{-n} (f_n^\pm)'(p_1p_{-1}) p_1 (-2p_0)\\
	&\stackrel{(\ref{derf})}{=} -(1-n)p_{n+1} + 2 \mu n p_0 p_{-1}^{-n-1} f_n^\pm(p_1p_{-1}) \pm \mu (n+1) p_0 p_1 p_{-1}^{-n} \frac{f_{n-1}^\pm(p_1p_{-1})}{\sqrt{\mu^2+p_1p_{-1}}} \\
	&\stackrel{\phi_0}{\simeq} -(1-n) p_{n+1} \mp 2\mu n \sqrt{\mu^2+p_1p_{-1}} p_{-1}^{-n-1} f_n^\pm (p_1 p_{-1}) - \mu (n+1) p_1 p_{-1}^{-n} f_{n-1}^\pm(p_1p_{-1})\\
	&\stackrel{\phi_{n+1}}{\simeq} (1-n)\mu p_{-1}^{-n-1} f_{n+1}(p_1p_{-1})\\
	&\quad\quad\quad \mp 2\mu n \sqrt{\mu^2+p_1p_{-1}} p_{-1}^{-n-1} f_n^\pm (p_1 p_{-1}) - \mu (n+1) p_1 p_{-1}^{-n} f_{n-1}^\pm(p_1p_{-1})\\
	&= -\mu p_{-1}^{-n-1} \left(
	(n-1)f_{n+1}(x) \pm 2 n \sqrt{\mu^2+x} f_n^\pm (x) + (n+1) x f_{n-1}^\pm(x)
	\right)_{x=p_1p_{-1}} 
	\stackrel{(\ref{rec3f})}{=}0,
\end{align*}
and
\begin{align*}
	\{{\bar\phi}_n,K_+\} &= \{p_{-n}+ \mu p_{-1}^n g_n^\pm(p_1p_{-1}),K_+ \} \\
	& = -(1+n)p_{-n+1} + \mu g_n^\pm (p_1p_{-1}) \{p_{-1}^n, K_+\} + \mu p_{-1}^n \{ g_n^\pm(p_1p_{-1}),K_+ \}\\
	& = -(1+n)p_{-n+1} + n\mu g_n^\pm(p_1p_{-1}) p_{-1}^{n-1} (-2p_0) + \mu p_{-1}^n (g_n^\pm)'(p_1p_{-1}) p_1 (-2p_0) \\
	& \stackrel{(\ref{derg}),\phi_0}{\simeq} -(1+n)p_{-n+1} \mp 2\mu n \sqrt{\mu^2 + p_1p_{-1}} p_{-1}^{n-1} g_n^\pm(p_1p_{-1}) + \mu (n-1) p_1 p_{-1}^n g_{n+1}^\pm(p_1p_{-1}) \\
	& \stackrel{\phi_{n-1}}{\simeq} (1+n)\mu p_{-1}^{n-1} g_{n-1}(p_1p_{-1}) \\
	&\quad\quad\quad \mp 2\mu n \sqrt{\mu^2 + p_1p_{-1}} p_{-1}^{n-1} g_n^\pm(p_1p_{-1}) + \mu (n-1) p_1 p_{-1}^n g_{n+1}^\pm(p_1p_{-1})\\
	&= \mu p_{-1}^{n-1} \left(
	(n+1)g_{n-1}(x) \pm 2 n \sqrt{\mu^2+x} g_n^\pm (x) + (n-1) x g_{n+1}^\pm(x)
	\right)_{x=p_1p_{-1}}
	\stackrel{(\ref{rec3g})}{=}0.
\end{align*}
Similarly, one can show that $\{\phi_n,K_-\}\simeq 0$, $ \{{\bar\phi}_n,K_-\}\simeq 0$. This can be done by direct calculation as above or using that the Poisson bracket structure is real and then
\begin{align}
	\{\phi_n,K_-\}^* = \{\bar{\phi}_n,K_+\} \simeq 0,\\
	\{{\bar\phi}_n,K_-\}^* =  \{{\phi}_n,K_+\} \simeq 0.
\end{align}

One concludes then that the extended Poincar\'e generators are conserved charges for our system.  A discussion of the  Casimirs of the Lorentz and Poincar\'{e} groups in BMS space is presented in Appendix \ref{secC}, where, in particular, it is shown that the only quadratic Casimir of the  BMS group is the standard one 
Poincar\'{e} Casimir, $C_2=p_0^2-p_1p_{-1}$, see appendix D.

One might be tempted to generalize the Lorentz generators to include superrotations (or rather superboosts) by replacing the ``1'' which appear in (\ref{K+}) and $(\ref{K-})$ with arbitrary positive integers $m$,
\begin{align}
	K_+^m &= \sum_{n=-\infty}^{+\infty} (m-n)x^n p_{n+m},\label{Km+}\\
	K_-^m &= \sum_{n=-\infty}^{+\infty} (m+n)x^n p_{n-m},\label{Km-}
\end{align}
for $m=1,2,\ldots$
By appropriate identifications, these generators, together with $J$ and the $\CP_n$, provide a representation of the extended BMS algebra in terms of Poisson brackets. However, the extended generators obtained for $m=2,3,\ldots$ do not commute with all the first class constraints of our system, and hence are not conserved quantities. To be more precise, the constraints $\phi_n$, $n=2,3,\ldots$, are weakly invariant only under $K_+^m$, while the $\bar{\phi}_n$ are invariant only under $K_-^m$. We will see in Section \ref{secM} that the massless limit of our theory is fully invariant under these generalized transformations.

\section{Massless limit}
\label{secM}
Since the Lagrangian (\ref{BMS_lag}) is proportional to $\mu$ one cannot take the massless limit directly in configuration space. This is not a problem in phase space, since the system is in this case entirely defined by the set of constraints, which have a non-trivial limit when $\mu\to 0$. Indeed, performing this limit in (\ref{rphi1}) and its complex conjugate (\ref{rphi2}) one gets the constraints
\begin{align}
	\varphi_n &= p_n \pm (\mp1)^n p_{-1}^{-n} (\sqrt{p_1p_{-1}})^{n+1},\quad n=0,1,2,\ldots,\label{rphi3}\\
	\bar{\varphi}_n &= p_{-n} \pm (\mp1)^n p_{-1}^n (\sqrt{p_1p_{-1}})^{-n+1},\quad n=1,2,\ldots,\label{rphi4}.
\end{align}
Notice that $\varphi_0=p_0\pm\sqrt{p_1p_{-1}}$, and that $\varphi_1$ and $\bar{\varphi}_1$ are trivial, as in the massive case.

As shown in Appendix \ref{appM}, one has that
\begin{equation}
	\{\varphi_n,K_+^m\} \simeq 0,\ 	\{\varphi_n,K_-^m\} \simeq 0,\
		\{\bar{\varphi}_n,K_+^m\} \simeq 0,\ 	\{\bar{\varphi}_n,K_-^m\} \simeq 0,
		\label{SRvarphi}
\end{equation}
where the superrotation generators $K_{\pm}^m$, $m=0,1,2,\ldots$ are defined as in (\ref{Km+}) and (\ref{Km-}), and where the weak equality is now over the manifold defined by $\varphi_n=0$, $\bar{\varphi}_n=0$. Thus, $K_{\pm}^m$ are conserved quantities in the massless limit theory.

The superrotation operators obey the algebra
\begin{align}
	\{K_+^m,K_+^n\} &= (m-n)K_+^{m+n},\\
	\{K_+^m,K_-^n\} &= \begin{cases}
	                	2mJ & \text{if $m=n$},\\
	               	-(m+n)K_+^{m-n} & \text{if $m>n$},\\
		                (m+n) K_-^{n-m} & \text{if $m<n$},
                    \end{cases}
                     \\
    \{K_-^m,K_-^n\} &= (m-n)K_-^{m+n},
\end{align}
Furthermore, they act on the supertranslation generators as
\begin{align}
\{K_\pm^m,p_n\} = (m\mp n) p_{n\pm m}, \quad m=1,2,\ldots,\quad n\in\Z.
\end{align}

If we now define
\begin{equation}
	L_m = \begin{cases}
		-J & \text{if $m=0$},\\
		K_+^m & \text{if $m>0$},\\
		-K_-^{-m} & \text{if $m<0$},
	\end{cases}
\end{equation}
it turns out that $\{L_m,L_n\} = (m-n) L_{m+n}$ and the extended BMS algebra (\ref{fullBMS}) is obtained in terms of Poisson brackets of the massless limit BMS particle.

\section{Gauge fixing}
\label{secGF}
In Section \ref{secCA} it has been shown that, after eliminating the degrees of freedom $u,v$ and its corresponding canonical momenta by means of the primary second-class constraints $\pi_u=0$, $\pi_v=0$, $\phi_1=0$ and $\bar{\phi}_1=0$, the theory still contains an infinite number of primary first-class constraints which generate gauge transformations and indicate the presence of gauge degrees of freedom. 

These gauge degrees can be eliminated by converting the first-class constraints to second class, by introducing appropriate gauge fixing conditions. Since all the constraints $\phi_n$ (resp.\ $\bar{\phi}_n$) depend linearly on $p_n$ (resp.\ $p_{-n}$), a sensible choice is to introduce the constraints
\begin{equation}
	\psi_n=x^n,\quad n=\pm2,\pm3,\ldots,
	\label{GF1}
\end{equation}
so that
\begin{equation}
	\{\psi_n,\phi_m\} = \delta_{n,m}, \quad n,m=\pm2,\pm3,\ldots,
	\label{GF10}
\end{equation}
and define a gauge fixing as
\begin{equation}
	GF=\{\psi_m=0\}_{|m|\geq 2},
	\label{GF2}
\end{equation}
which allows the consistent elimination of all the extra BMS degrees of freedom in phase space,
\begin{align}
	p_{\pm n} = -\frac{\mu}{p_{\mp1}^n} f_n^\pm(p_1p_{-1}), \quad x^n=0, \quad x^{-n}=0,\quad n=2,3,\ldots,
	\label{gf1}
\end{align}
with only the gauge symmetry associated with $\phi_0$ remaining. In this way, the physical degrees of freedom of the theory in phase space are reduced to $x^0, p_0, x^{\pm1}, p_{\pm1}$, and it can be seen that the Dirac brackets between these remaining variables are the standard Poisson brackets.

Notice that the gauge condition is not invariant under supertranslations and hence one must introduce a compensating gauge transformation so that the total variation of the gauge condition, computed on the gauge condition, is zero. If we consider $|m|\geq 2$ and denote by $\delta_{ST}^n x^m$ the supertranslation of $x^m$ generated by  $p_n$, and by $\epsilon^m(\tau)$ the gauge transformation on $x^m$, generated by $\phi_m$, one has
\begin{equation}
	0 = \left.\left( \epsilon^m(\tau) + \delta_{ST}^n x^m\right)\right|_{GF},
\end{equation}
and, since $\delta_{ST}^n x^m= \epsilon^n \delta_n^m=\epsilon^m$, the compensating gauge transformation associated to the supertranslation along the $m$ coordinate, $|m|\geq 2$, is just
\begin{equation}
	\epsilon^m(\tau) = -\epsilon^m.
	\label{CSTm}
\end{equation}

Since the generators of the gauge transformations $\phi_m$, $|m|\geq 2$, contain the momenta $p_1, p_{-1}$, it turns out that these compensating gauge transformations induce a residual transformation on the remaining variables $x^{\pm 1}$, given by
\begin{equation}
	\delta^m_{\text{res}} x^{\pm 1} = \{ x^{\pm 1}, -\epsilon^m \phi_m  \},\quad |m|\geq 2.
\end{equation}
Using $\{x^1,\phi_n\} = -(n+1)P_{n-1}/(2P_0)$, $\{x^{-1},\phi_n\} = (n-1)P_{n+1}/(2P_0)$, $n=\pm2,\pm3,\ldots$, one gets 
\begin{align}
	\delta^m_{\text{res}} x^{1} &= \epsilon^m (m+1) \frac{P_{m-1}}{2P_0},\label{resGT1}\\
		\delta^m_{\text{res}} x^{-1} &= -\epsilon^m (m-1) \frac{P_{m+1}}{2P_0},
		\label{resGT2}
\end{align}
where it should be reminded that the several $P_n$ appearing on the right-hand sides are  functions of $p_1$, $p_{-1}$. That these transformations are a symmetry of the theory is proved at the end of Appendix \ref{appVL}. Notice that for $m=1$ and $m=-1$, although no compensating gauge transformation is needed, one formally obtains the standard translations in $x^1$ and $x^{-1}$, respectively.

These residual transformations on the physical variables $x^{\pm1}$ provide, together with the Lorentz transformations, a realization of BMS in the physical reduced space, up to reparametrizations. The need for a reparametrization follows from the fact that $x^0$ does not transform under $\delta^m_{\text{res}}$ but, under a boost, transforms into $x^{1}$ or $x^{-1}$. For instance, for $K_+$ one has $\delta_+ x^0=\{x^0,K_+\}=2x^{-1}$ and then (we drop the constant parameters $\epsilon^m$)
\begin{equation}
	[\delta_+,\delta^m_{\text{res}}]x^0 =(m-1)\frac{P_{m+1}}{P_0},
\end{equation}
while $\delta^{m+1}_{\text{res}}x^0$, which should appear on the right-hand side in order to have the BMS algebra, is zero. However, 
since
 $\{x^0,\phi_0\}=1$, 
the right-hand side can be interpreted as a reparametrization with parameter
\begin{equation}
	\epsilon_+^m = (m-1)\frac{P_{m+1}}{P_0}, \label{eq:parameter}
\end{equation}
so that the commutator is indeed a vanishing BMS supertranslation on $x^0$ plus a reparametrization,
\begin{equation}
	[\delta_+,\delta^m_{\text{res}}]x^0 = (m-1)\cdot 0 + \delta_0^{\epsilon_+^m} x^0.
	\label{resK+0}
\end{equation}
For this to be consistent, the same reparametrization should appear when one considers the action of the transformations on $x^1$ and $x^{-1}$,
\begin{align}
	[\delta_+,\delta^m_{\text{res}}] x^{1} &=  \frac{m+1}{2} \left( \frac{p_1 P_{m-1}}{P_0^2} + m \frac{P_{m-2}}{P_0}  \right),\label{resK+1}\\
	[\delta_+,\delta^m_{\text{res}}] x^{-1} &= - \frac{m-1}{2} \left( \frac{p_1 P_{m+1}}{P_0^2} + m \frac{P_{m+2}}{P_0}  \right).\label{resK+2}
\end{align}
Using that
\begin{align}
	\delta^{m+1}_{\text{res}} x^1 = (m+2)\frac{P_m}{2P_0},& \quad \delta^{m+1}_{\text{res}} x^{-1} = -m\frac{P_{m+2}}{2P_0},
\end{align}
Notice this residual transformation is no longer a point transformation.
Under reparametrizations with parameter $\epsilon_+^m$ \eqref{eq:parameter},
\begin{align}
	\delta_0^{\epsilon_+^m} x^1 &= \epsilon_+^m \{x^1,\phi_0\} = -\frac{m-1}{2}\frac{p_{-1}P_{m+1}}{P_0^2},\\
	\delta_0^{\epsilon_+^m} x^{-1} &= \epsilon_+^m \{x^{-1},\phi_0\} = -\frac{m-1}{2}\frac{p_{1}P_{m+1}}{P_0^2},
\end{align}
one can check that (\ref{resK+1}), (\ref{resK+2}) can be rewritten as
\begin{align}
	[\delta_+,\delta^m_{\text{res}}] x^{1} &=  (m-1)\delta^{m+1}_{\text{res}} x^1 + \delta_0^{\epsilon_+^m} x^1 ,\\
	[\delta_+,\delta^m_{\text{res}}] x^{-1} &= (m-1)\delta^{m+1}_{\text{res}} x^{-1} + \delta_0^{\epsilon_+^m} x^{-1},
\end{align}
which, together with (\ref{resK+0}) and up to the reparametrization, yield the correct term for the BMS algebra. Similarly, for $K_-$, the reparametrization parameter is
\begin{equation}
	\epsilon_-^m = -(m+1)\frac{P_{m-1}}{P_0},
\end{equation}
while no reparametrization is necessary to close the commutators of (\ref{resGT1},\ref{resGT2}) with the transformation given by the rotation generator $J$.

This is an example of a fact previously reported in the literature (see \cite{Pons:1994hy}, eq.~(3.11)), \textit{i.e.} the closure of the algebra of rigid symmetries with the help of gauge transformations. 
In our case, the rigid transformations correspond to Lorentz and supertranslations in physical space, and the gauge transformation is given by the reparametrization invariance associated with the first-class constraint $\phi_0$, which has not been fixed.
That the reparametrizations might be needed to close the algebra can be also be inferred from the fact that the constraints $\phi_n$, $|n|\geq 2$, are weakly Lorentz invariant on the manifold defined by $\phi_0$ (see the proof in Section \ref{secGBMS}) and that those $\phi_n$ are the generators of the residual gauge transformations that give rise to the BMS symmetry in physical space.

Under a Lorentz transformation,  
\begin{align}
	\delta_J x^n = \{x^n,J\} = n x^n,\\
	\delta_+ x^n = \{x^n,K_+\} = (2-n) x^{n-1},\\
	\delta_- x^n = \{x^n,K_-\} = (2+n) x^{n+1},
\end{align}
and one has
\begin{align}
	\left.\delta_J x^n\right|_{GF} = 0,\quad
	\left.\delta_+ x^n\right|_{GF} = 0,\quad
	\left.\delta_- x^n\right|_{GF} = 0,
\end{align}
so that the gauge condition is preserved, without the need for a compensating gauge transformation. Notice that the factors $(2-n)$ and $(2+n)$ play a fundamental role for $n=2$ and $n=-2$, respectively.

In the massless case, where the Lorentz group can be extended so as to include superrotations, one has, for $m=2,3,\ldots$,
\begin{align}
	\delta_+^m x^n = \{x^n,K_+^m\} = (2m-n) x^{n-m},\label{GF11a}\\
	\delta_-^m x^n = \{x^n,K_-^m\} = (2m+n) x^{n+m},\label{GF11b}
\end{align}
and a compensating gauge transformation must be introduced for $|n|\geq 2$ for the values of $m$ such that the right-hand side of (\ref{GF11a}) or (\ref{GF11b}) are not zero when evaluated on the gauge fixing condition (\ref{GF2}). As in the case of supertranslations, this will generate a residual gauge transformation for $x^{\pm 1}$, which should provide a realization of superrotations.

In any case, after eliminating the gauge degrees of freedom, the remaining variables are just $x^0$, $x^{\pm 1}$ and their canonical momenta, together with the first class constraint $\phi_0$. This describes a Poincar\'e particle in $2+1$, with the corresponding reparametrization invariance, with no extra degrees of freedom, and with a realization of the supertranslations, plus superrotations in the massless case, provided by the residual gauge transformations. Summing up, the physical degrees of freedom of the BMS particle do not contain the BMS coordinates for $|n|>1$ and hence no soft BMS modes are present.

\section{Conclusions and outlook}
\label{secCon}

We have constructed a non-linear realization of a massive particle Lagrangian for the BMS symmetry algebra in $2+1$ space-time. This Lagrangian depends on an infinite set of BMS coordinates, which include the standard $2+1$ Poincar\'e ones, together with the Goldstone boost variables. 

The canonical analysis of this Lagrangian reveals the existence of a finite set of second-class constraints, which can be eliminated using the standard Dirac bracket construction, together with an infinite set of first-class constraints, which generate a corresponding infinite set of gauge transformations. 

The standard Lorentz generators in $2+1$ are extended so that they act on the full set of BMS variables, and the theory is shown to be invariant under them. These extended Lorentz generators can be further generalized to an infinite set, the so-called superrotations, that obey the extended $\text{BMS}_3$ algebra, but it is only in the massless limit of the theory that they are conserved quantities and thus represent a symmetry of the system.

Upon fixing all the gauge degrees of freedom of the theory, except for the standard reparametrization, one obtains a theory whose physical content is that of an ordinary relativistic Poincar\'e particle, with the standard reparametrization invariance provided by the remaining first-class constraint. However, this gauge fixing procedure results in residual gauge transformations acting on the standard space coordinates $x^0, x^1$, $x^{-1}$ which, modulo reparametrizations, realize the BMS symmetry. Since the remaining first-class constraint is the standard one, the field theory associated with this particle model is that of a free Klein-Gordon field.

The interpretation of these transformations for $x^1$, $x^{-1}$, which depend on the associated canonical momenta $p_1$, $p_{-1}$, is a subject for further study. The appearance of the momenta in a non-polynomial form seems to indicate that, in the field theory associated with this model, the field should transform non-locally. In \cite{Batlle:2017llu,Batlle:2022hwf} a realization of BMS in terms of a free scalar field was constructed, with transformations that were non-local in ordinary space. Further investigations are needed to clarify whether the two approaches can be related.

	A possible link with the nonlocal transformations for the massless scalar field presented in \cite{Batlle:2022hwf} is as follows.
	
Since the standard field equations can be obtained from the particle mass-shell condition following the world-line approach to field theory, one can consider, for the massless BMS particle, the infinite tower of squares of the constraints 
	\begin{equation}
		\varphi_n\bar{\varphi}_n \simeq p_np_{-n} - p_1 p_{-1}, \ n=0,2,3,\ldots
		\label{gKG1}
	\end{equation}
	(the case $n=1$ is trivial). For $n=0$, $p_0^2-p_1p_{-1}=0$, this yields the standard massless Klein-Gordon equation. Indeed, using the relation with the real coordinates $p_{\pm1}= P_1\mp i P_2$ and $p_0=-i\partial_{x^0}$, $P_j = -i\partial_{x_j}$, $j=1,2$, 
	one has
	\begin{equation}
		(p_0^2-p_1p_{-1})\Psi = (-\partial^2_{x^0}+\nabla^2)\Psi =0.
		\label{KG}
	\end{equation} 
	For $n=2,3\ldots$, one can introduce similar real coordinates $x^{(n)}_{1,2}$
	related to the complex BMS coordinates $x^{\pm n}$, $n\geq 2$, by
	\begin{equation}
		x^{\pm n} = \frac{1}{2} (x^{(n)}_1\pm i x^{(n)}_2),\quad n=2,3,\ldots
	\end{equation}
	so that
	\begin{equation}
		p_{{\pm n}}=P^{(n)}_1\mp P^{(n)}_2 = -i \frac{\partial}{\partial x^{(n)}_1}\mp \frac{\partial}{\partial x^{(n)}_2}.
	\end{equation}
	 The imposition of all the conditions on an scalar field $\Psi(x^0,x_1,x_2,x^{(2)}_1,x^{(2)}_2,\ldots)$, depending on all of the BMS coordinates, leads then to the infinite set of equations
	\begin{equation}
		(p_np_{-n}-p_1p_{-1})\Psi = (-\nabla_n^2+\nabla^2)\Psi =0,\quad \nabla_n^2 =  \frac{\partial^2}{\partial x^{(n)2}_1}+\frac{\partial^2}{\partial x^{(n)2}_2}, \ n=2,3,\ldots
	\end{equation} 
	together with (\ref{KG}). 
	
	Notice that $ p_np_{-n} - p_1 p_{-1}=0$, $n=0,2,3,\ldots$, can be written in the equivalent form
	\begin{equation}
		p_np_{-n} - p_{n+1} p_{-(n+1)}=0, \ n=0,1,2,3,\ldots,
		\label{gKG2}
	\end{equation}
	which, upon converted to a field equation, yields the standard Klein-Gordon equation plus an infinite tower of equations,
	\begin{equation}
		\nabla^2_{n+1}\Psi = \nabla^2_{n}\Psi, \quad n=1,2,\ldots,
		\label{gKG3}
	\end{equation}
or, equivalently, the infinite set of Klein-Gordon equations
\begin{equation}
	\nabla^2_{n}\Psi = \partial_{x^0}^2\Psi, \quad n=1,2,\ldots
	\label{gK4}
\end{equation}
Whether the Lie symmetries of this set of PDE can be related to the nonlocal symmetries of an scalar field depending only on the ordinary Poincar\'e variables $x^0, x^{(1)}$, which were  obtained in 
\cite{Batlle:2022hwf} in terms of polyharmonic functions, is a subject worthy of further study.

In the approach taken in this paper, the massless limit has been obtained in the Hamiltonian formalism as the theory defined by the massless limit of the Hamiltonian constraints. An alternative approach, which we will examine in the future, is to obtain the model of a massless particle in the nonlinear realization approach.

The construction of a particle model exhibiting BMS symmetry presented in this paper could be, in principle, repeated for $\BMS_4$, using, for instance, the stereographic parametrization of $\BMS_4$ \cite{Barnich:2009se,Barnich:2011ct}. BMS structure constants for $\BMS_4$, $BMS_5$ and $BMS_6$ using generalized spherical harmonics parametrizations can also be found in  \cite{Delmastro:2017erq}, although they are much more involved.

\acknowledgments
We acknowledge interesting discussions with Luca Ciambelli, Miguel Campliglia, Jaume Gomis, Marc Henneaux, Axel Kleinschmidt and Sabrina Pasterski.
JG acknowledges the hospitality of the Max Planck Albert Einstein Institute in Golm and the Perimeter Institute in Waterloo where this work has been completed.
The work of CB is partially supported by Project  MASHED (TED2021-129927B-I00), funded by MCIN/AEI/10.13039/501100011033 and by the European Union Next Generation EU/PRTR.  JG has been supported in part by
PID2019-105614GB-C21 and PID2019- 105614GB-C21 and from the State Agency for Research of the Spanish Ministry of Science and Innovation 
through the Unit of Excellence Maria de Maeztu 2020-2023 award to the Institute of Cosmos Sciences (CEX2019-000918-M).

\appendix
\section{Computation of the term of the Maurer-Cartan form proportional to the $P_0$ generator}
\label{appA}

Using 
\begin{equation}
	e^{X}Y e^{-X} = e^{\operatorname{ad} _X} Y =Y+\left[X,Y\right]+\frac{1}{2!}[X,[X,Y]]+\frac{1}{3!}[X,[X,[X,Y]]]+\cdots
\end{equation} 
one has
\begin{eqnarray}
	\left[Y, e^{-X}\right] &=& e^{-X} \left( \left[X,Y\right]+\frac{1}{2!}[X,[X,Y]]+\frac{1}{3!}[X,[X,[X,Y]]]+\cdots \right) 
	\nonumber\\
	&\equiv& e^{-X} K(X,Y),
	\label{derBCH}
\end{eqnarray}
where
\begin{equation}
	K(X,Y) = \left[X,Y\right]+\frac{1}{2!}[X,[X,Y]]+\frac{1}{3!}[X,[X,[X,Y]]]+\cdots,
\end{equation}
which is linear in its second argument. We will call $K(X,Y)$ the $K$-action of $X$ on $Y$.

By repeated use of (\ref{derBCH}) one arrives at
\begin{equation}
	U^{-1}P_n U = P_n + K(-iL_1 u,P_n) + K(-iL_{-1} v,P_n) + K(-i L_1 u, K(-i L_{-1}v,P_n)).
	\label{iUPU}
\end{equation}

The  second and third terms in (\ref{iUPU})  are

\begin{eqnarray}
	K(-iL_1 u,P_n) &=& \sum_{l=1}^\infty \frac{1}{l!} (-iu)^l [L_1,[L_1, \stackrel{l)}{\ldots},[  L_1,P_n ]    \ldots               ] ] \nonumber\\
	&=& \sum_{l=1}^\infty \frac{1}{l!} (-iu)^l (-i)^l (n-1)n\ldots (n+l-2) P_{n+l}\nonumber\\
	&=& \sum_{l=1}^\infty \frac{1}{l!}  (-1)^l u^l  (n-1)n\ldots (n+l-2) P_{n+l}.\label{LpPn}
\end{eqnarray}

\begin{eqnarray}
	K(-iL_{-1} v,P_n) &=& \sum_{l=1}^\infty \frac{1}{l!} (-iv)^l [L_{-1},[L_{-1}, \stackrel{l)}{\ldots},[  L_{-1},P_n ]    \ldots               ] ] \nonumber\\
	&=& \sum_{l=1}^\infty \frac{1}{l!} (-iv)^l (-i)^l (n+1)n\ldots (n-l+2) P_{n-l}\nonumber\\
	&=& \sum_{l=1}^\infty \frac{1}{l!}  (-1)^l v^l  (n+1)n\ldots (n-l+2) P_{n-l}.\label{LmPn}
\end{eqnarray}

The fourth term in (\ref{iUPU}) is

\begin{eqnarray}
	K(-iL_1 u, K(-i L_{-1}v,P_n)) &=& K(-iL_1 u,
	\sum_{l=1}^\infty \frac{1}{l!}  (-1)^l v^l  (n+1)n\ldots (n-l+2) P_{n-l}
	)\nonumber\\
	&=& 
	\sum_{l=1}^\infty \frac{1}{l!}  (-1)^l v^l  (n+1)n\ldots (n-l+2) K(-iL_1 u,P_{n-l})\nonumber, 
\end{eqnarray}
where we have used the linearity of $K$ in its second argument. Finally
\begin{eqnarray}
	\lefteqn{ K(-iL_1 u, K(-i L_{-1}v,P_n))=\sum_{l=1}^\infty \frac{1}{l!}  (-1)^l v^l  (n+1)n\ldots (n-l+2)} 
	\nonumber \\
	& & 
	\sum_{k=1}^\infty \frac{1}{k!}(-1)^k u^k (n-l-1)(n-l)\ldots (n-l+k-2) P_{n-l+k}.
	\label{LpLmPn}
\end{eqnarray}

As mentioned before, we are interested only in the $P_0$ terms in (\ref{iUPU}):

\begin{itemize}
	\item $P_n$. Contributes only for $n=0$, with $P_0$.
	\item  $K(-iL_1 u,P_n)$. Only the terms with $l=-n$ yield a $P_0$. Since $l\geq 1$, this means that there is no contribution for $n\geq 0$, while for $n=-m<0$ one picks the term
	$$
	\frac{1}{m!} (-1)^m u^m (-m-1)(-m)\ldots (-2) P_0  = (m+1)u^m P_0.
	$$
	\item $K(-iL_{-1} v,P_n)$. The $P_0$ contribution is obtained now for $l=n$ and since $l\geq 1$, there is only contribution if $n>0$, which is
	$$
	\frac{1}{n!} (-1)^n v^n (n+1)n\ldots 2\, P_0 = (-1)^n (n+1) v^n P_0.
	$$

	\item $K(-iL_1 u, K(-i L_{-1}v,P_n))$. This term has  multiple $P_0$ contributions, given by $l-k=n$, subjected to $l\geq 1$, $k\geq 1$. For given $l$ one picks the $k=l-n$ term in the $k$ series, but $k\geq 1$ implies that $l$ must satisfy, besides $l\geq 1$, the constraint $l\geq 1+n$. If $n\leq 0$ this just means $l\geq 1$, but, for $n>0$, $l$ is restricted by $l\geq 1+n$. 
	
	Selecting $k=l-n$ in (\ref{LpLmPn}) and restricting the series over $l$ according to the above discussion one has, after re-arranging terms and cancelling  some signs,
	\begin{equation}
		\sum_{l=n+1}^\infty \frac{l-n+1}{l!} (-1)^l  (n+1)n \ldots (n-l+2) v^l u^{l-n} P_0
	\end{equation}
	for $n>0$, and
	\begin{equation}
		\sum_{l=1}^\infty \frac{m+l+1}{l!}   (m-1)m \ldots (m+l-2) v^l u^{l+m} P_0
	\end{equation}
	for $n=-m\leq 0$.
\end{itemize}

Putting everything together, the coefficient of $P_0$ in (\ref{MCg}) can be computed as follows, collecting the contributions proportional to the different $\dif x^n$.

For $n=0$, there is only contribution from the first and fourth terms in (\ref{iUPU}),
$$
\dif x^0 (1+ \sum_{l=1}^\infty \frac{l+1}{l!} (0-1)(0)\ldots (0+l-2) v^l u^l
$$	
Notice, however, that the above series finishes in fact after $l=1$, so one gets
\begin{equation}
	\dif x^0 (1-2uv).
\end{equation}	

For $n>0$, only the third and fourth terms have a non-vanishing contribution, given by	
\begin{equation}
	\dif x^n \left(  
	(-1)^n (n+1) v^n +  \sum_{l=n+1}^\infty \frac{l-n+1}{l!} (-1)^l  (n+1)n \ldots (n-l+2) v^l u^{l-n}
	\right).
\end{equation}
Actually, the product in the coefficients of the series always contains a zero except if $l=n+1$, and hence the above expression collapses to
\begin{equation}
	\dif x^n \left(  
	(-1)^n (n+1) v^n +  2 (-1)^{n+1} u v^{n+1}
	\right)= \dif x^n  (-1)^n v^n (n+1- 2 uv).
\end{equation}

Finally, for $n=-m<0$, the contributions come from the second and fourth terms and are given by	
\begin{equation}
	\dif x^{-m} \left(  
	(m+1) u^m +   \sum_{l=1}^\infty \frac{m+l+1}{l!}   (m-1)m \ldots (m+l-2) v^l u^{l+m}
	\right).
\end{equation}
The series is identically zero for $m=1$, while for $m\geq 2$
it can be rewritten as
\begin{equation}
	\dif x^{-m} \left(  
	(m+1) u^m +   \sum_{l=1}^\infty \frac{m+l+1}{l!}   \frac{(l+m-2)!}{(m-2)!} v^l u^{l+m}
	\right).
\end{equation}
This series can be summed (provided that $|uv|<1$)
  and, after adding the $(m+1)u^m$ term, one gets
\begin{equation}
	\dif x^{-m}  u^m  \frac{m+1-2uv}{(1-uv)^{m}}.
\end{equation}

Adding all the terms, the coefficient of $P_0$ in the Maurer-Cartan form is

\begin{equation}
\Omega_{P_{0}}=	\dif x^0 (1-2uv)
	+ \sum_{n=1}^\infty  \dif x^n  
	(-1)^n v^n (n+1- 2 uv)
	+ \dif x^{-1} 2u+    \sum_{n=2}^\infty 
	\dif x^{-n} 
	u^n  \frac{n+1-2uv}{(1-uv)^{n}}.
	\label{BMS_P00}
\end{equation}

The fact that the $\dot x^{-n}$ contribution is much more complex than that of $\dot x^n$, actually involving the series that has been mentioned, is due to the form of the last term in (\ref{iUPU}), which in turn is a consequence of the ordering that we have selected for the two exponentials in $U$. For $n\geq 2$, the $K$-action of $-i vL_{-1}$ on $P_n$ can only descend to $P_{-1}$ (since the Poincar\'e part is BMS invariant), and then there is only one term in the $K$-action of $-iL_1 u$ that returns to $P_0$. Instead, for $n\geq 2$,
$K(-i vL_{-1}, P_{-n})$ produces terms $P_k$ for any $k=-3,-4,\ldots$, and then, for each of them, there is a  way   to return to  $P_0 $  by the 
$K$-action of $-iL_1 u$.

\section{Quasi-invariance of the Lagrangian under gauge transformations}
\label{appVL}
We consider first the full Lagrangian (\ref{BMS_lag}) and its variation under the full set of gauge transformations given by the first class constraints $\phi_m$ and  $\bar{\phi}_m$, $m\geq 2$.
In order to compute the transformation of the phase-space variables induced by $\phi_m$, $\bar{\phi}_m$ one needs
\begin{align}
	\{x^1,\phi_m\} &= \{x^1,p_m+\mu p_{-1}^{-m} f_m^\pm(p_1p_{-1})\} \nonumber\\
	&= \mu p_{-1}^{-m} (f_{m}^\pm)'(p_1p_{-1}) p_{-1} = \mp \mu p_{-1}^{-m+1} \frac{m+1}{2\sqrt{\mu^2+p_1p_{-1}}} f_{m-1}^\pm(p_1p_{-1}) \nonumber\\
	&= \mu p_{-1}^{-m+1} \frac{m+1}{2 P_0} f_{m-1}^\pm(p_1p_{-1})
	= -(m+1) \frac{P_{m-1}}{2P_0},\\ 
	\{x^1,\bar{\phi}_m\} &= \{x^1,\bar{p}_m+\mu p_{-1}^{m} g_m^\pm(p_1p_{-1})\} \nonumber\\
	&=\mu p_{-1}^m (g_{m}^\pm)'(p_1p_{-1}) p_{-1} 
	=\pm\mu p_{-1}^{m+1}\frac{m-1}{2\sqrt{\mu^2+p_1p_{-1}}}g_{m+1}^\pm(p_1p_{-1})
	\nonumber\\
	&= -\mu p_{-1}^{m+1} \frac{m-1}{2P_0} g_{m+1}^\pm(p_1p_{-1}) 
   = (m-1)\frac{P_{-m-1}}{2P_0}.
\end{align}
From these two it also follows that
\begin{align}
	\{x^{-1},\phi_m\} &= \{x^{1},\bar{\phi}_m\}^* = (m-1) \frac{P_{m+1}}{2P_0},\\
	\{x^{-1},\bar{\phi}_m\} &= \{x^{1},{\phi}_m\}^* = -(m+1) \frac{P_{-m+1}}{2P_0},
\end{align}
which, together with $\{x^{n},\phi_m\}=\delta^n_m$, $\{x^{-n},\bar{\phi}_m\}=\delta^n_m$, $\{P_n,\phi_m\}=\{P_{-n},\phi_m\}=\{P_n,\bar{\phi}_m\}=\{P_{-n},\bar{\phi}_m\}=0$, allow to compute the transformations of all the terms in the Lagrangian. For instance, if $G=\alpha_m\phi_m=\epsilon_m 2P_0\phi_m$, one has
\begin{align}
	\delta{\cal L} &= \frac{\dif}{\dif\tau} (2P_0\epsilon_m) P_m + \frac{\dif}{\dif\tau}  (-(m+1)\epsilon_m P_{m-1})p_1+ \frac{\dif}{\dif\tau}  ((m-1)\epsilon_m P_{m+1})p_{-1}\nonumber\\
     &= \epsilon_m (2\dot{P}_0 p_m-(m+1)\dot{P}_{m-1}p_1+(m-1)\dot{P}_{m+1} p_{-1}) \nonumber\\
     &\quad + \dot{\epsilon}_m (2P_0P_m-(m+1)p_1P_{m-1}+(m-1)P_{m+1} p_{-1}).
     	\label{eqVL2}
\end{align}
This can be written as a total derivative, $\delta{\cal L}=\frac{\dif}{\dif\tau}F$, provided that
\begin{equation}
	2P_0\dot{P}_m - (m+1)P_{m-1}\dot{p}_1+(m-1)P_{m+1}\dot{p}_{-1} = 0.
	\label{eqVL3}
\end{equation}
 Using\footnote{We do not display the dependence of $f_m^\pm$ on $p_1p_{-1}$.}
\begin{align}
	\dot{P}_m &= \frac{\dif}{\dif\tau}\left(-\frac{\mu}{p_{-1}^m} f_m^\pm\right)
	= m \frac{\mu}{p_{-1}^{m+1}}\dot{p}_{-1} f_m^\pm - \frac{\mu}{p_{-1}^m} (f_m^\pm)'\cdot (\dot{p}_1 p_{-1}+p_1 \dot{p}_{-1})\\
	&= m \frac{\mu}{p_{-1}^{m+1}}\dot{p}_{-1} f_m^\pm - \frac{\mu}{p_{-1}^m} \left(
	\mp \frac{m+1}{2\sqrt{\mu^2 + p_1p_{-1}}} f_{m-1}^\pm
	\right) (\dot{p}_1 p_{-1}+p_1 \dot{p}_{-1}),
\end{align}
the left-hand side of (\ref{eqVL3}) is 
\begin{align}
\text{LHS(\ref{eqVL3})} &= \mp 2 \sqrt{\mu^2 + p_1p_{-1}}m \frac{\mu}{p_{-1}^{m+1}}\dot{p}_{-1} f_m^\pm -(m+1)\frac{\mu}{p_{-1}^m} f_{m-1}^\pm \cdot (\dot{p}_1 p_{-1}+p_1 \dot{p}_{-1})\nonumber\\
&\quad - (m+1)P_{m-1}\dot{p}_1+(m-1)P_{m+1}\dot{p}_{-1}\nonumber\\
 &= \mp 2 \sqrt{\mu^2 + p_1p_{-1}}m \frac{\mu}{p_{-1}^{m+1}}\dot{p}_{-1} f_m^\pm -(m+1)\frac{\mu}{p_{-1}^m} f_{m-1}^\pm \cdot (\dot{p}_1 p_{-1}+p_1 \dot{p}_{-1})\nonumber\\
&\quad +(m+1)\frac{\mu}{p_{-1}^{m-1}} f_{m-1}^\pm\dot{p}_1-(m-1)\frac{\mu}{p_{-1}^{m+1}}f_{m+1}^\pm\dot{p}_{-1}
\end{align}
The two terms containing $\dot{p}_1$ cancel each other, while the terms proportional to $\dot{p}_{-1}$ are
\begin{align}
-\dot{p}_{-1}\frac{\mu}{p_{-1}^{m+1}}\left(
(m-1)f_{m+1}^\pm \pm 2 m \sqrt{\mu^2 + p_1p_{-1}} f_m^\pm  + (m+1) p_1p_{-1} f_{m-1}^\pm
\right),
\end{align}
which is zero due to (\ref{rec3f}). This proves (\ref{eqVL3}) and thus (\ref{VLphi}). Equation (\ref{VLbphi}) is proved in a similar way.

We consider next the partially gauge fixed Lagrangian
\begin{equation}
	{\cal L}_0 = \dot{x}^0 P_0 + \dot{x}^1 p_1 + \dot{x}^{-1} p_{-1},
\end{equation}
obtained from the full Lagrangian by setting $x^m=0$ for $|m|\geq 2$, and which, as explained in the text, is just the standard Lagrangian for a massive Poincaré particle. This Lagrangian has the gauge symmetry transformation induced by the remaining first-class constraint $\phi_0$, associated with reparametrization invariance, and the Poincar\'{e} invariance generated by $p_0$, $p_1$, $p_{-1}$, $J$, $K_+$ and $K_{-}$, but also the infinite set of symmetries given by the residual gauge transformations (\ref{resGT1}), (\ref{resGT2}), which we write in the condensed notation
\begin{align}
	\delta_{\text{res}}^m x^1 &= \epsilon^m (m+1) \frac{P_{m-1}}{2P_0} = A_m(p_1,p_{-1}),\\
	\delta_{\text{res}}^m x^{-1} &= -\epsilon^m (m-1) \frac{P_{m+1}}{2P_0} = B_m(p_1,p_{-1}),
\end{align} 
with all the other variables $x^0$, $p_0$, $p_1$ and $p_{-1}$ invariant. One has
\begin{align}
	\delta_{\text{res}}{\cal L}_0 &= \delta_{\text{res}}(\dot{x}^1 p_1 + \dot{x}^{-1} p_{-1}) =  p_1 \frac{\dif}{\dif\tau}A_m +  p_{-1} \frac{\dif}{\dif\tau}B_m
=  C_m \dot{p}_1 +  D_m \dot{p}_{-1},
\end{align}
where
\begin{align}
	C_m &= p_1 \frac{\partial A_m}{\partial p_1} + p_{-1} \frac{\partial B_m}{\partial p_1},\\
	D_m &= p_1 \frac{\partial A_m}{\partial p_{-1}} + p_{-1} \frac{\partial B_m}{\partial p_{-1}}.
\end{align}
The quasi-invariance of the Lagrangian, that is, the existence of a function $F_0$ such that $\delta_{\text{res}}{\cal L}_0=\frac{\dif}{\dif\tau}F_0$, is equivalent to
\begin{equation}
	\frac{\partial C_m}{\partial p_{-1}} = \frac{\partial D_m}{\partial p_{1}},
\end{equation}
which boils down to
\begin{equation}
	\frac{\partial A_m}{\partial p_{-1}} = \frac{\partial B_m}{\partial p_{1}},
\end{equation}
which in turn can be proved using the expressions of $P_n$, $P_0$ in terms of $p_1$ and $p_{-1}$ and the properties of the functions $f_n^\pm$.
 
\section{Invariance of the massless limit constraints under superrotations}
\label{appM}
Using $\{p_n,K_+^m\}=(n-m) p_{n+m}$ one has that the variation of $\varphi_n$ under a superrotation induced by $K_+^m$ is
\begin{align*}
\{\varphi_n,K_+^m\} &=\{p_n \pm (\mp1)^n p_{-1}^{-n} (\sqrt{p_1p_{-1}})^{n+1},K_+^m\} \\
&= (n-m)p_{n+m} \pm (\mp)^n (-n p_{-1}^{-n-1} (-1-m)p_{-1+m}) (\sqrt{p_1p_{-1}})^{n+1}\\
&\pm (\mp)^n p_{-1}^{-n}(n+1) (\sqrt{p_1p_{-1}})^{n}\frac{1}{2\sqrt{p_1p_{-1}}}
\left(
p_1 (-1-m)p_{-1+m}+p_{-1}(1-m)p_{1+m}
\right).
\end{align*}
Since we only have to deal with the case $m\geq 2$, one has that $n+m$, $-1+m$ and $1+m$ are all positive. Using $\varphi_{n+m}$, $\varphi_{-1+m}$ and $\varphi_{1+m}$ one can express   
$p_{n+m}$, $p_{-1+m}$ and $p_{1+m}$ in terms of $p_1$ and $p_{-1}$, and one obtains, after re-arranging terms and extracting the common dependency of all the terms in  $p_1$ and $p_{-1}$,
\begin{align*}
	\{\varphi_n,K_+^m\} &\simeq (\mp)^{n+m+1} p_{-1}^{-n-m}  (\sqrt{p_1p_{-1}})^{n+m+1}\\
	& \cdot \left(
	n-m - n (1+m) +(1+m)\frac{n+1}{2} - (1-m)\frac{n+1}{2}
	\right) =0.
\end{align*}

Similarly, from $\{p_n,K_-^m\}=-(m+n)p_{n-m}$,
\begin{align*}
	\{\varphi_n,K_-^m\} &=\{p_n \pm (\mp1)^n p_{-1}^{-n} (\sqrt{p_1p_{-1}})^{n+1},K_-^m\} \\
	&= -(n+m)p_{n-m} \pm (\mp)^n (-n p_{-1}^{-n-1}( -(m-1)p_{-1-m})) (\sqrt{p_1p_{-1}})^{n+1}\\
	&\pm (\mp)^n p_{-1}^{-n}(n+1) (\sqrt{p_1p_{-1}})^{n}\frac{1}{2\sqrt{p_1p_{-1}}}
	\left(
	p_1 (-(m-1)p_{-1-m})+p_{-1}(-(m+1)p_{1-m})
	\right).
\end{align*}
Again, since we must only consider $m\geq 2$, both $p_{-1-m}$ and $p_{1-m}$ are BMS momenta with negative indexes, and can be expressed in terms of $p_1$ and $p_{-1}$ using $\bar{\varphi}_{1+m}$ and $\bar{\varphi}_{m-1}$, respectively. For $n-m\geq 0$, one can use $\varphi_{n-m}$ for $p_{n-m}$, while for $n-m<0$ $p_{n-m}$ has negative index and can be expressed in terms of $p_1$ and $p_{-1}$ using $\bar{\varphi}_{m-n}$. It turns out that in both cases the term obtained from $p_{n-m}$ is the same, and one has
 \begin{align*}
 	\{\varphi_n,K_-^m\} &\simeq (\mp)^{n+m+1} p_{1}^{-m} p_{-1}^{-n}  (\sqrt{p_1p_{-1}})^{n+m+1}\\
 	& \cdot \left(
 	-(n+m) - n (m-1) +(m-1)\frac{n+1}{2} + (m+1)\frac{n+1}{2}
 	\right) =0. 
 \end{align*}
It should be noticed that it is this case, the variation of a positive index constraint under a negative index superrotation, the one that breaks the invariance of the theory under superrotations in the massive case.

Due to the real character of the Poisson bracket, one will also have
\begin{align*}
	\{\bar{\varphi}_n,K_-^m\}^* &= \{\varphi_n,K_+^m\} \simeq 0,\\
		\{\bar{\varphi}_n,K_+^m\}^* &= \{\varphi_n,K_-^m\} \simeq 0,
\end{align*}
and thus all the constraints are weakly invariant under all the superrotations.

\section{Casimirs of the Lorentz and Poincar\'e groups in BMS space}
\label{secC}
The action of the Lorentz generators $K_\pm$, $J$ on the $p_n$, $n\in\Z$, provided by Poisson brackets,
\begin{align}
	\delta_J p_n = \{p_n,J\} = -n p_n,\\
	\delta_+ p_n = \{p_n,K_+\} = -(1-n) p_{n+1},\\
	\delta_- p_n = \{p_n,K_-\} = -(1+n) p_{n-1},
\end{align}
leads to the definition of infinite dimensional matrices acting on vectors $(\ldots,p_{-2},p_{-1},p_0,p_1,p_2,\ldots)$ which implement this action, given by
\begin{align}
	J_{nm} &= - n \delta_{nm},\\
	(K_+)_{nm} &= - (1-m) \delta_{n,m+1},\\
	(K_-)_{nm} &= - (1+m) \delta_{n,m-1}.
\end{align}
Using the structure constants of the $SO(2,1)$ algebra in the $J,K_+,K_-$ basis one can construct the Killing form of the Lie algebra, and from that the quadratic Casimir, which is given by 
\begin{equation}
	C^L_2 = \frac{1}{2}J^2 + \frac{1}{4} K_+K_- +\frac{1}{4}K_-K_+.
	\label{C2}
\end{equation}
Using the above matrices one immediately obtains
\begin{equation}
	\left(\frac{1}{2}J^2 + \frac{1}{4} K_+K_- +\frac{1}{4}K_-K_+\right)_{nm} = 1\cdot \delta_{nm}
\end{equation}
and hence this corresponds to an adjoint representation on the space of BMS momenta.

Similarly, one can consider the action on the space of BMS coordinates $x^n$, $n\in\Z$,
\begin{align}
	\delta_J x^n = \{x^n,J\} = n x^n,\\
	\delta_+ x^n = \{x^n,K_+\} = (2-n) x^{n-1},\\
	\delta_- x^n = \{x^n,K_-\} = (2+n) x^{n+1},
\end{align}
which leads to the matrices 
\begin{align}
	\tilde{J}_{nm} &= n \delta_{nm},\\
	(\tilde{K}_+)_{nm} &= (2-m) \delta_{n,m-1},\\
	(\tilde{K}_-)_{nm} &=  (2+m) \delta_{n,m+1}.
\end{align}
Again
\begin{equation}
	\left(\frac{1}{2}\tilde{J}^2 + \frac{1}{4} \tilde{K}_+\tilde{K}_- +\frac{1}{4}\tilde{K}_-\tilde{K}_+\right)_{nm} = 1\cdot \delta_{nm},
\end{equation}
which shows that it also corresponds to an adjoint representation on the space of coordinates.

With respect to the Poincar\'e group, the quadratic Casimir in $2+1$ in our coordinates is
\begin{equation}
	C_2^P = p_0^2-p_1p_{-1},
	\label{C2P}
\end{equation}
which, for our system and taking into account the constraint $\phi_0$, takes value $-\mu^2$. One may wonder if it is possible to obtain a quadratic Casimir involving the higher BMS momenta, of the form
\begin{equation}
	C_2 = A_{mn}p_mp_n,\quad A_{mn}=A_{nm}.
\end{equation}
Imposing the invariance under $J$ one gets
\begin{equation}
	\delta_J C_2 = - A_{mn} p_m p_n (m+n) =0
\end{equation}
which implies that the only $A_{mn}$ that can be different from zero are those corresponding to $m=-n$. Thus
\begin{equation}
	A_{mn} = A_n \delta_{m,-n},	
\end{equation}
with $A_n=A_{-n}$ due to the symmetry of $A_{mn}$. Computing now the variation under $K_+$ and using this form for $A_{mn}$ one gets
\begin{equation}
	\delta_+ C_2 = - p_{m+1}p_{-m} ((1-m)A_m+(2+m)A_{m+1}) =0.
\end{equation}
In order to equal to zero the coefficients of this sum over $m$ one must notice that the terms corresponding to $m=n$ and $m=-1-n$ yield the same product  $p_{m+1}p_{-m}$. Taking this into account and using  $A_m=A_{-m}$ one gets the first order recurrence relation
\begin{equation}
	(1-m) A_m + (2+m) A_{m+1} =0, \quad m=0,1,2,\ldots.
	\label{recA}
\end{equation}
The invariance under $K_-$ does not add any new condition.
For $m=0$, (\ref{recA}) yields
$$
A_0+2 A_1=0,
$$ 
from which $A_1=-\frac{1}{2}A_0$ and hence also $A_{-1}=-\frac{1}{2}A_0$. For $m=1$, however, the relation is
$$
0\cdot A_1 + 3 A_2 =0,
$$
from which $A_2=0$ and thus $A_{-2}=0$. From this point, using the recurrence for higher values of $m$ leads to $A_m=A_{-m}=0$ for $m=2,3,\ldots$. The final result is then that the only quadratic Casimir of the Poincar\'e group in BMS space is, up to a global constant, the standard one, given by (\ref{C2P}).

\bibliographystyle{jhep}
\bibliography{bms}

\end{document}